\documentclass{jaa}
\usepackage{natbib}
\usepackage{graphicx}
\usepackage{ulem}

\begin{document}\sloppy

%%paper title
%%For line breaks \\ can be used within title
%\title{Title of the paper goes here:\\ Second line}
\title{TIFR Treasures for Astronomy from Ground to Space}

%%author names are separated by comma (,)
%%use \and before the last author name
%%use a * along with the number separated by comma
%% for the  author for correspondence
%%\textsuperscript{number} is used for affiliation
%%\affilOne, \affilTwo etc., upto \affilTwentyfive is possible
%%Please note the first letter after \affil is capitalised in the command
%%

\author{Supriyo Ghosh\textsuperscript{1}, Devendra K. Ojha\textsuperscript{1,*}, Saurabh Sharma\textsuperscript{2} and Milind B. Naik\textsuperscript{1}}
\affilOne{\textsuperscript{1}Tata Institute of Fundamental Research, Homi Bhabha Road, Colaba, Mumbai 400 005, India.\\}
\affilTwo{\textsuperscript{2}Aryabhatta Research Institute of Observational Sciences, Manora Peak, Nainital-263001, Uttarakhand, India.}

%%escape two column mode for title, affiliation and abstract
%%by giving \twocolumn command as shown

\twocolumn[{

\maketitle
%%include \corres to print the corresponding author Email id
\corres{ojha@tifr.res.in}

%%include \msinfo for
%%manuscript information such as
%%received, revised and accepted dates
%%
\msinfo{...}{...}

%%abstract
\begin{abstract}
The infrared astronomy group of Department of Astronomy and Astrophysics at Tata Institute of Fundamental Research has been pursuing astronomical instrumentation activities since its inception. The group has been routinely involved in balloon-borne astronomy program from field station at Hyderabad with indigenously developed payloads. Ground-based astronomical activities began with a single element infrared detector. Later, over time, larger format array detectors are being used in the cameras. These astronomy cameras have been routinely used at observatories across India. Recently, the group has also developed a laboratory model of the Infrared Spectroscopic Imaging Survey payload, targeted for the small satellite mission of the Indian Space Research Organisation, which will carry out spectroscopic measurements in the wavelength range 1.7 to 6.4 $\mu$m seamlessly.
\end{abstract}

%%insert keywords separated by 3 hyphens using \keywords{words}

}]
%%close the twocolumn escape here

%%include \doinum{number}for the DOI number in the header
%%include \volnum{number} for the volume number in the header
%%include \year{yyyy} for  year of publication in the header
%%include \pgrange{num--num} page range of article in the header
%%include \artcitid{num} for the article citation id
%%include \lp to print last page of the article
%%include \setcounter{page}{pagenum} for the exact starting page of the article

\doinum{12.3456/s78910-011-012-3}
\artcitid{\#\#\#\#}
\volnum{000}
\year{2021}
\pgrange{1--10}
\setcounter{page}{1}
\lp{1}

\section{Introduction}
The research themes and interests of the infrared astronomy (IRA) group at Tata Institute of Fundamental Research (TIFR) are primarily focused on the study of the interstellar medium (ISM) concerning star formation in our Galaxy and nearby galaxies. The dust and gas-filled ISM predominantly emits in the infrared waveband. Thus, the study of infrared emitting ISM helps us to probe several important astrophysical phenomena, namely, star formation activity, shock front, material recycling, photo processes in the proximity of young stars, and plasma cooling \citep{2018BSRSL..87...58O}. Observational investigations are carried out using TIFR’s own ground-based near-infrared (NIR) imagers and spectrometers and also other imagers and spectrometers available on national optical telescopes, TIFR’s indigenously developed 100-cm balloon-borne far-infrared (FIR) telescope, the Giant Metrewave Radio Telescope as well as international facilities such as the large aperture ground-based telescopes and astronomical satellites. The interpretation of observational data is done in the light of image processing as well as numerical modelling with the help of codes developed in-house as well as those available publicly. In addition, the IRA group is an active contributor in instrument development for ground-based and space-based astronomy.

It was way back in the 1940s that the ballooning activity in TIFR was started for atmospheric exploration and cosmic ray studies. Initially, hydrogen-filled rubber balloons were used to send the scientific equipment to altitudes of 25--30 km. In 1959, a balloon flight took place for the first time using TIFR own manufactured polyethene balloon. Over the years, the size of the payload became progressively larger and heavier with the complexity of experiments \citep{2021arXiv210311803A}. In the mid-80s, astronomy programmes were also started using the 100-cm balloon-borne FIR telescope in the FIR band (120 to 300 $\mu$m) \citep{1988ApJ...330..928G, 2010ASInC...1..167G}. Recently, as a part of the TIFR-Japan collaboration in balloon-borne FIR astronomy, a Japanese Fabry-Perot Spectrometer (FPS) tuned to the astrophysically interesting [C {\footnotesize II}] fine structure line at 157.74 $\mu$m line was installed at the focal plane of the 100 cm TIFR balloon-borne FIR telescope (T100), and subsequently, several flights took place to map large regions in [C {\footnotesize II}] line and continuum of several northern and southern star-forming complexes \citep{2001BASI...29..337M, 2003A&A...404..569M, 2013A&A...556A..92K, 2021A&A...651A..30S}. We refer to this combination of instruments (FPS + T100) as FPS100.

To complement these studies in the NIR bands, the need was felt for dedicated NIR instruments, which could be used with available 2$-$4 m class telescopes in India. In addition, with the development of NIR detector materials and arrays, NIR observations became on par with optical observations. More importantly, the NIR regime is very important to study
of many astrophysically cool objects like evolved giants and super-giants, low-mass stars, red and brown dwarfs. Hence, the IRA group aims to develop their own NIR instruments to meet this need. The first outcome was the development of the NIR imaging camera, TIFR Near Infrared Imaging Camera - I (TIRCAM1). It consisted of a 58 $\times$ 62 pixels InSb focal plane array (FPA), sensitive to 1--5 $\mu$m. It was mounted at the f/13 Cassegrain focus of the Mount Abu 1.2 m telescope belonging to Physical Research Laboratory, India, and was operational during the period 2001$-$2006. Additional details of the camera can be found in \citet{1993BASI...21..485G}, \citet{2005BASI...33..133G} and \citet{2002BASI...30..827O, 2003BASI...31..467O, 2006MNRAS.368..825O}. Later in 2012, the camera optics was re-designed and optimized for observations with the 2 m Himalayan Chandra Telescope (HCT; f/9) operated by the Indian Institute of Astrophysics, India, and the 2 m Girawali Telescope (f/10) operated by the Inter-University Centre for Astronomy and Astrophysics, India. The camera was also upgraded with a larger format detector array, Raytheon 512 $\times$ 512 pixels InSb Aladdin III Quadrant FPA to utilize its full capability in the range of 1 to 3.7 $\mu$m. The camera was therefore renamed as TIFR Near Infrared Imaging Camera-II (TIRCAM2). Later, the instrument has been tested by the IRA group at the 3.6 m Devasthal Optical Telescope (DOT), at an altitude of 2424 m above mean sea level and is being used by Indian and Belgian astronomers since May 2017. In addition, the IRA group designed and built TIFR Near Infrared Spectrometer and Imager (TIRSPEC) in collaboration with M/s. Mauna Kea Infrared LLC, Hawaii, USA (hereafter MKIR) during the 2007-2012 five-year period, to provide spectrometry in the 1.0 to 2.5 $\mu$m band with a medium spectral resolving power of $\sim$1200 \citep{2012ASInC...4..191O, 2014JAI.....350006}. It is now in operation on the side port of the 2 m HCT, Hanle (Ladakh), India, at an altitude of 4500 m above mean sea level. Furthermore, for near-simultaneous observations in the optical and NIR bands, the IRA group developed a dedicated Optical-NIR spectrometer, TIFR-ARIES Near Infrared Spectrometer (TANSPEC) for DOT in collaboration with Aryabhatta Research Institute of Observational Sciences (ARIES) and MKIR. It covers an exceptionally wide spectral range from 0.5 to 2.5 $\mu$m in a single exposure, with an intermediate spectral resolving power of $\sim$ 2750 \citep{2018BSRSL..87...58O}. It is now operational from the axial port of DOT.

In addition, the need was felt for a dedicated survey instrument in the wavelength regime 1.7--6.4 $\mu$m, which could be used for an Indian Satellite. To meet this need, The Infra-Red Spectroscopic Imaging Survey (IRSIS) payload was conceived to carry out an infrared spectroscopic imaging survey in the wavelength regime from 1.7 to 6.4 $\mu$m \citep{2010ASInC...1..171G}.

In this paper, we describe the technical details and current status of TIRCAM2 (Section 2), TIRSPEC (Section 3), TANSPEC (Section 4), FPS100 (Section 5) and IRSIS (Section 6). We describe TIFR ground-based instruments in greater detail and provide preliminary results on characterization and performance of TIRCAM2 on the side port and TANSPEC on the axial port of DOT.

\section{TIFR Near Infrared Imaging Camera-II (TIRCAM2)} \label{TIRCAM2}
The TIRCAM2 is an in-house instrument of the TIFR-IRA group. It is a closed cycle cooled NIR imager having wavelength coverage ranging from 1 to 3.7 $\mu$m. The TIRCAM2 employs a 512$\times$512 InSb based Aladdin III Quadrant FPA that is cooled to 35 K during operation by a closed-cycle Helium cryo-cooler for optimizing the quantum efficiency (which drops at lower temperatures) and the dark current (which increases with higher temperature). The primary motivation to develop the camera was to map Polycyclic Aromatic Hydrocarbons (PAH) emission at 3.3 $\mu$m in Galactic star-forming regions (GSFRs), however, the most distinctive feature of this camera is the capability of carrying out observation in the narrow-band L (nbL at 3.59 $\mu$m) band. The TIRCAM2 is the only NIR imaging camera currently available in India which is efficient to observe up to the L band. In addition to PAH and nbL filters,  J, H, Kcont, K and Br$\gamma$ are the other available standard filters in the instrument for imaging observations. The camera was initially mounted on the 2 m Girawali telescope and then installed on the axial port of the DOT telescope. The TIRCAM2 is now operational from one of the side ports of DOT in parallel with any instrument on the axial port of DOT. Additional technical details of the imager can be found in \citet{naik2012tircam2} and \citet{2012ASInC...4..189O}.

\subsection{TIRCAM2 on the axial port of DOT}
The TIRCAM2 was installed on the axial port of DOT on June 1, 2016 and the first light was observed on June 2, 2016.  The performance tests of the TIRCAM2 camera with the DOT were carried out during  May 11--14, 2017 and October 7--31, 2017. The TIRCAM2 covers a field of view (FoV) of 86.5 arcsec $\times$ 86.5 arcsec with an image scale of 0.169 arcsec/pixel on the DOT. During the calibration night, the seeing was typically in sub-arcsecond in spite of relatively high humidity ($>$ 70\%) with the best of $\sim$ 0.45 arcsec on October 16, 2017, in the K band. It was found that the 10$\sigma$ limiting magnitudes are 19 mag, 18.8 mag and 18 mag in J, H and K bands with corresponding exposures of 550 s, 550 s and 1000 s, respectively, under typical seeing conditions. The camera has also the capability to observe sources up to 9.2 mag (detection limit) in the nbL band with a net exposure of 20 s. Thus, the camera can be used as a good complementary instrument to carry out observation of the sources (brighter than $\sim$ 9.2 mag) that are saturated in the  Spitzer-Infrared Array Camera 3.6 $\mu$m and the Wide-field Infrared Survey Explorer W1 (at 3.4 $\mu$m) images. In addition, the camera was found to be very efficient in detecting strong PAH emitting sources (F$_{3.3 \mu m}$  $\geq$ 0.4 Jy), like Sh 2-61. Additional details of the performance on the axial port of DOT can be found elsewhere (\citealt{2018BSRSL..87...58O, 2018cwla.conf..257O, doi:10.1142/S2251171718500034}). To summarize, the TIRCAM2 on the axial port of DOT performed as expected, the camera is competent to other worldwide available instruments on 4 m class telescopes for deep NIR observations, and the results are highly encouraging at longer wavelengths, specifically at the nbL band at 3.59 $\mu$m. It was made available to the Indian and Belgian astronomical community for scientific observations since May 2017. A number of important observations were carried out successfully since then (see, e.g., \citealt{2020JApA...41...27A, 10.1093/mnras/staa2412, 10.1093/mnras/staa2403, 2021arXiv210912365A}). However, due to the availability of larger instruments, for instance, TANSPEC, ARIES-Devasthal Faint Object Spectrograph \& Camera, and CCD imager, which can only be mounted on the axial port of DOT, it was decided to try out the TIRCAM2 camera at one of the side ports of DOT.

%The typical seeing during the calibration nights was in sub-arcsecond at the telescope site instead of relatively high humidity ($>$ 70\%.), and the best seeing of $\sim$ 0.45 arcsec was obtained in the K band on October 16, 2017. Deep imaging observations show that we can observe sources up to 19 mag, 18.8 mag and 18 mag (S/N $\sim$ 10) in the J, H and K band with corresponding exposures of 550 s, 550 s and 1000 s, respectively, and up to 9.2 mag (detection limit) in the nbL band with a net exposure of 20 s under typical seeing conditions.
\subsection{TIRCAM2 on the side port of DOT}
\begin{figure*}
    \centering\includegraphics[height=.3\textheight]{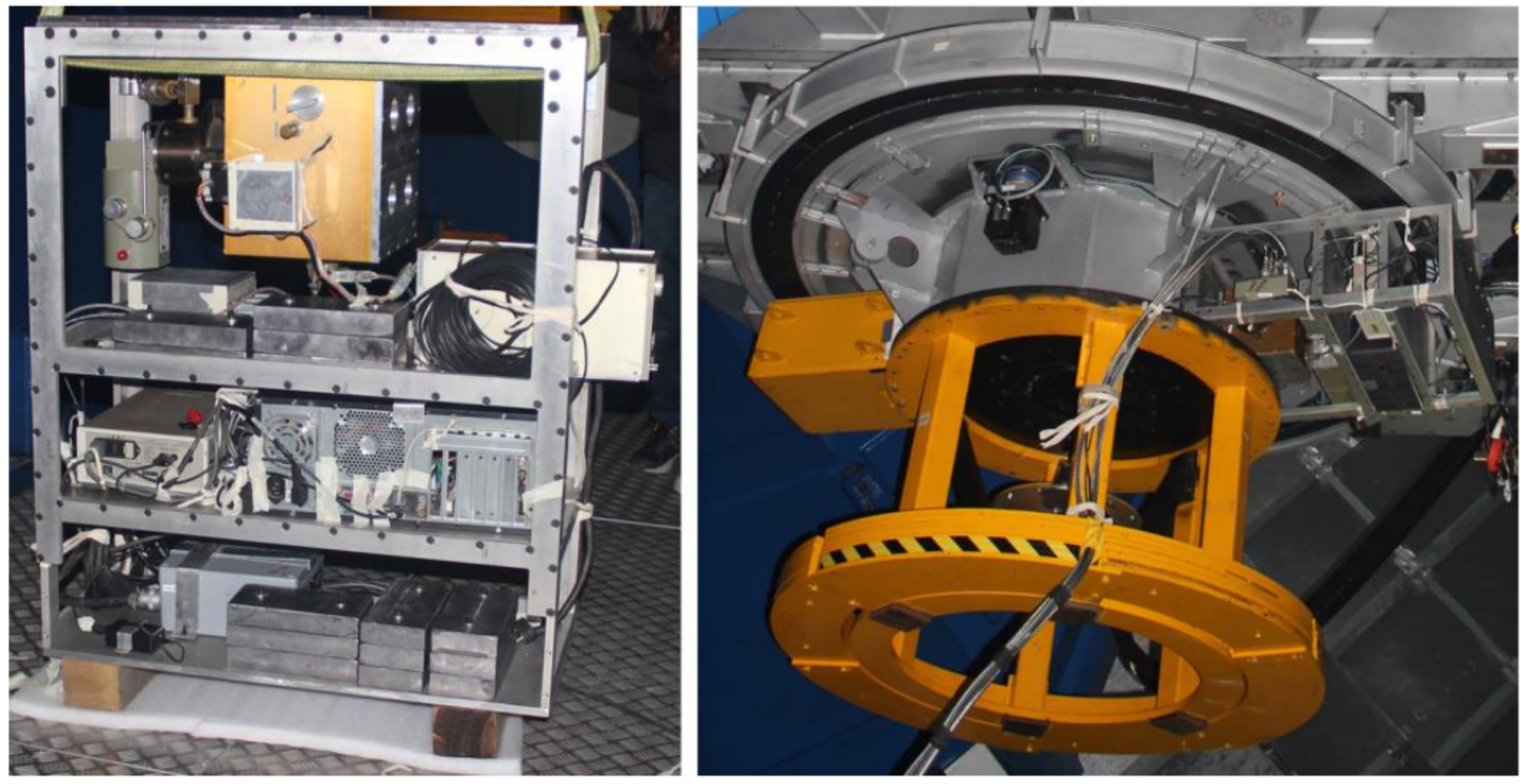}
    \caption{Complete TIRCAM2 rack assembly before mounting (left) and mounted rack on the side port (right) of DOT.}
    \label{Fig:TIRCAM2_DOT-side-port}
\end{figure*}

 To install the TIRCAM2 instrument on the side port of the telescope, the design of the mechanical interface needs to be changed. However, it was very challenging considering the size of the instrument interface (which is primarily meant for the axial port) and the envelop size of the side port. The instrument interface plate and rack for mounting the camera and related electronics equipment have been successfully redesigned at TIFR central workshop, and the TIRCAM2 camera was mounted on the side port of the DOT telescope in May 2020 in collaboration with ARIES, Nainital as shown in Fig.~\ref{Fig:TIRCAM2_DOT-side-port}. Because of the permanent mount of the TIRCAM2 camera on the side port, one can easily switch over the axial port instrument to TIRCAM2 almost immediately. This is a major milestone of DOT scientific research because the efficiency of the telescope will drastically increase with two instruments being available for near-simultaneous observations during any particular night, and one can obtain near simultaneous observations in optical and NIR. The first light images of TIRCAM2 on the side port were taken on May 4, 2020. It was astounding that in the first light we were able to observe science images. However, during the first engineering run, a few short exposure TIRCAM2 images were only possible to observe for testing the camera on the side port because of the bad weather. The performance of TIRCAM2 on the side port of DOT was tested based on these short exposure images.
 
\begin{figure*}
	\centering\includegraphics[height=.495\textheight]{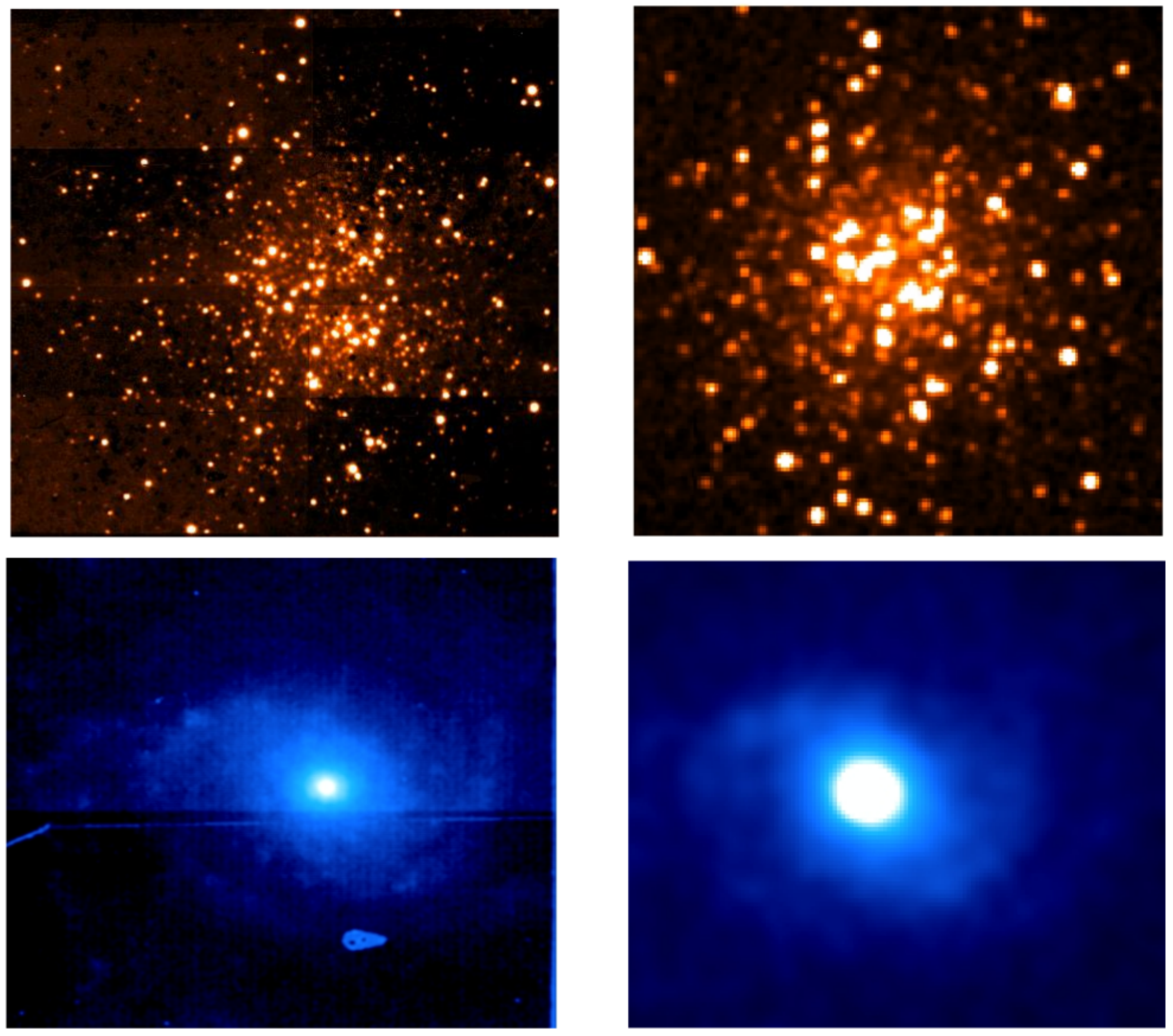}	 
	\caption{Top panels: A mosaic J band TIRCAM2 image (FoV $\sim$ 2.6 arcmin $\times$ 2.6 arcmin) of a Galactic globular cluster, M53 observed with the exposure time of 50 s is shown (left). The 2MASS J band image of a similar FoV is also shown (right) for comparison. The TIRCAM2 image had a FWHM of $\sim$ 0.8 arcsec. Bottom panels: The J band TIRCAM2 image (exposure time of 50 s, FoV $\sim$ 86.5 $\times$ 86.5 square arcsec) of M99 galaxy is displayed (left). The 2MASS image of the galaxy is also shown (right) for comparison. It can be clearly seen in the figures that TIRCAM2 images are much deeper than the 2MASS images.}
	\label{Fig:Comaparsion_between_2massVsTIRCAM2}
\end{figure*} 

\begin{figure}
	\centering\includegraphics[width=7.5cm,height=5cm]{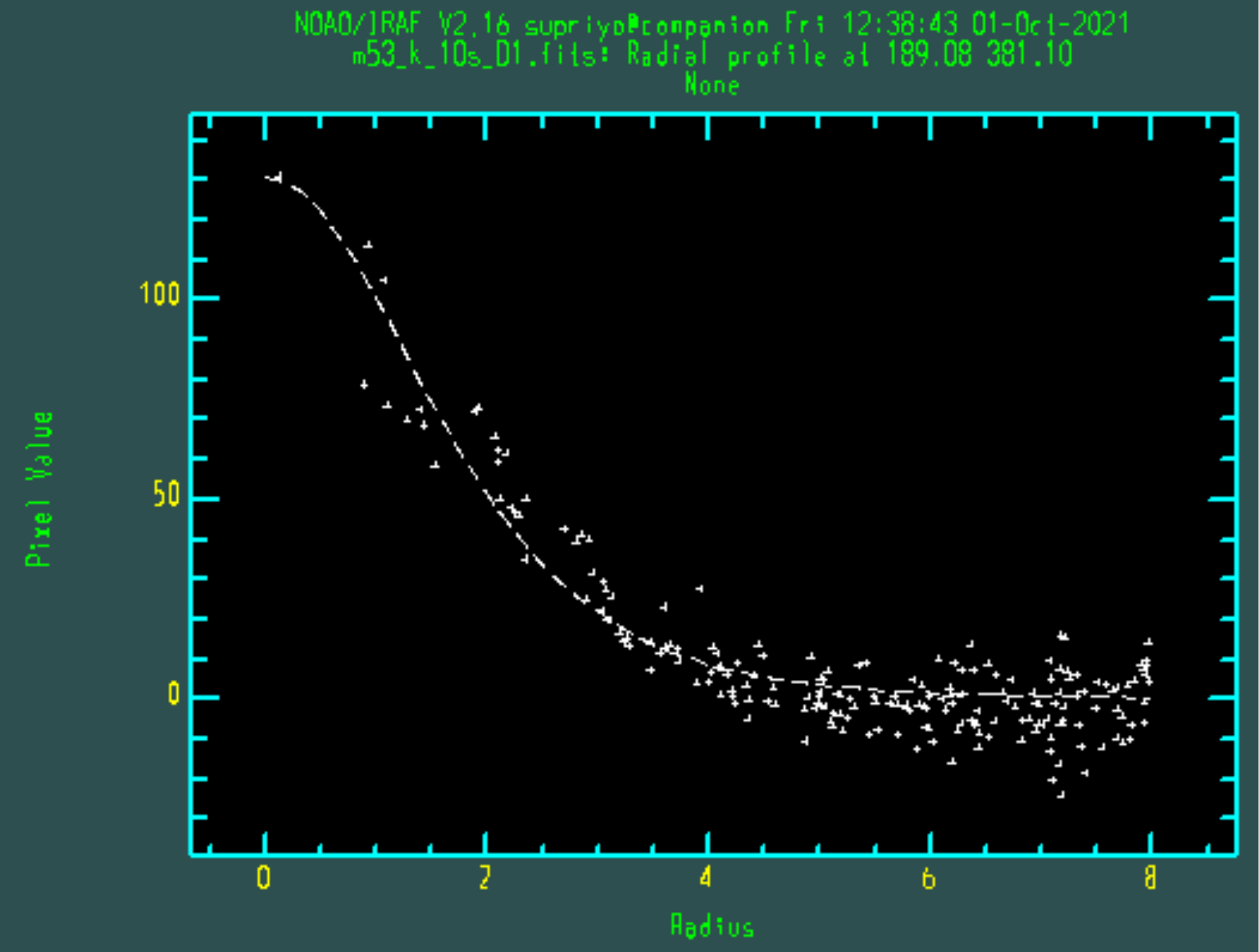}	\\
	\caption{Radial profile of a star observed with the TIRCAM2 on the side port of DOT in the K band on May 4, 2020. The FWHM of the profile is $\sim$ 4.0 pixels which converts to $\sim$ 0.7 arcsec on the sky.}
	\label{Fig: radial_profile_TIRCAM2}
\end{figure} 

\begin{figure}
	\centering\includegraphics[width=8cm,height=6cm]{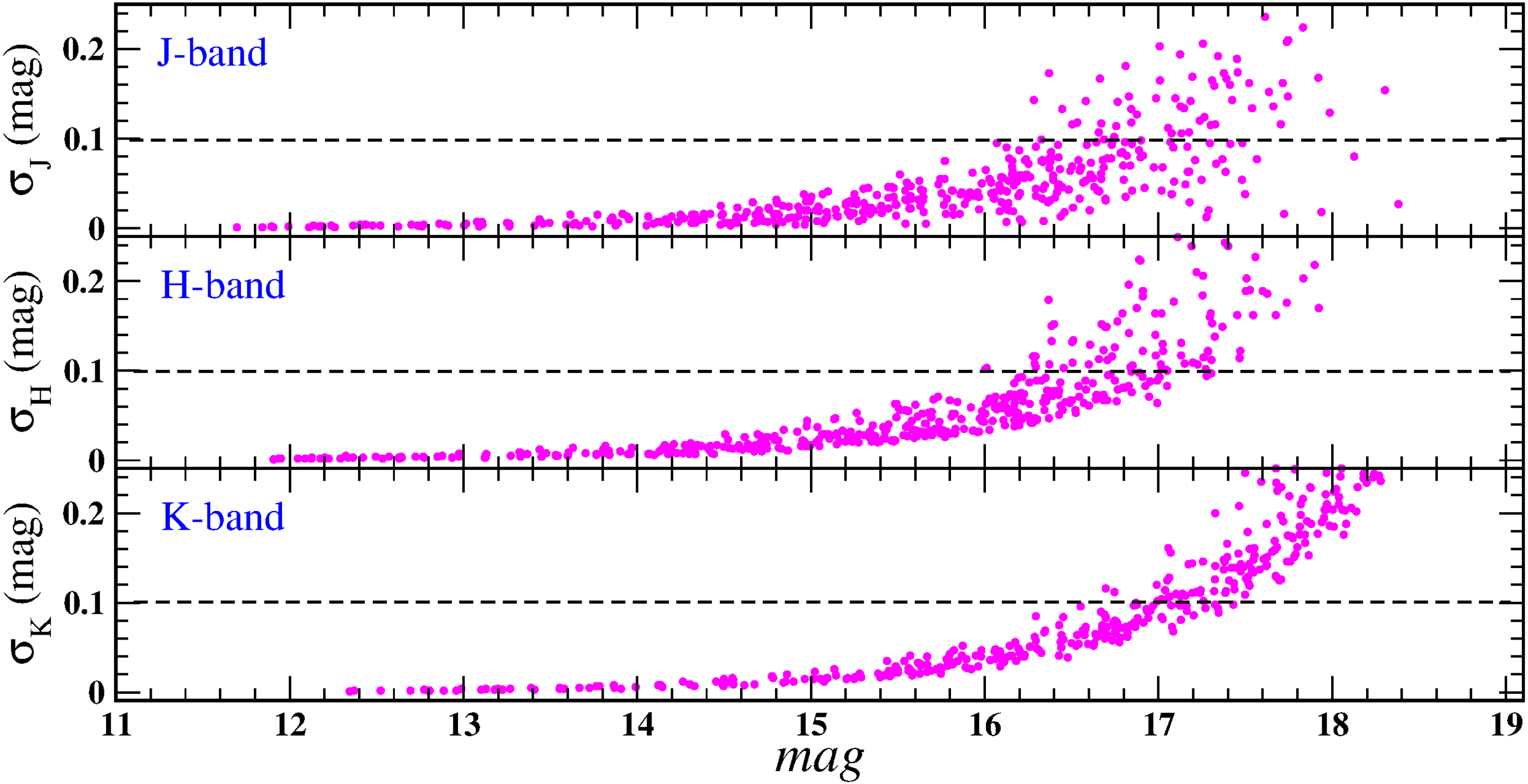}	\\
	\centering\includegraphics[width=8cm,height=6cm]{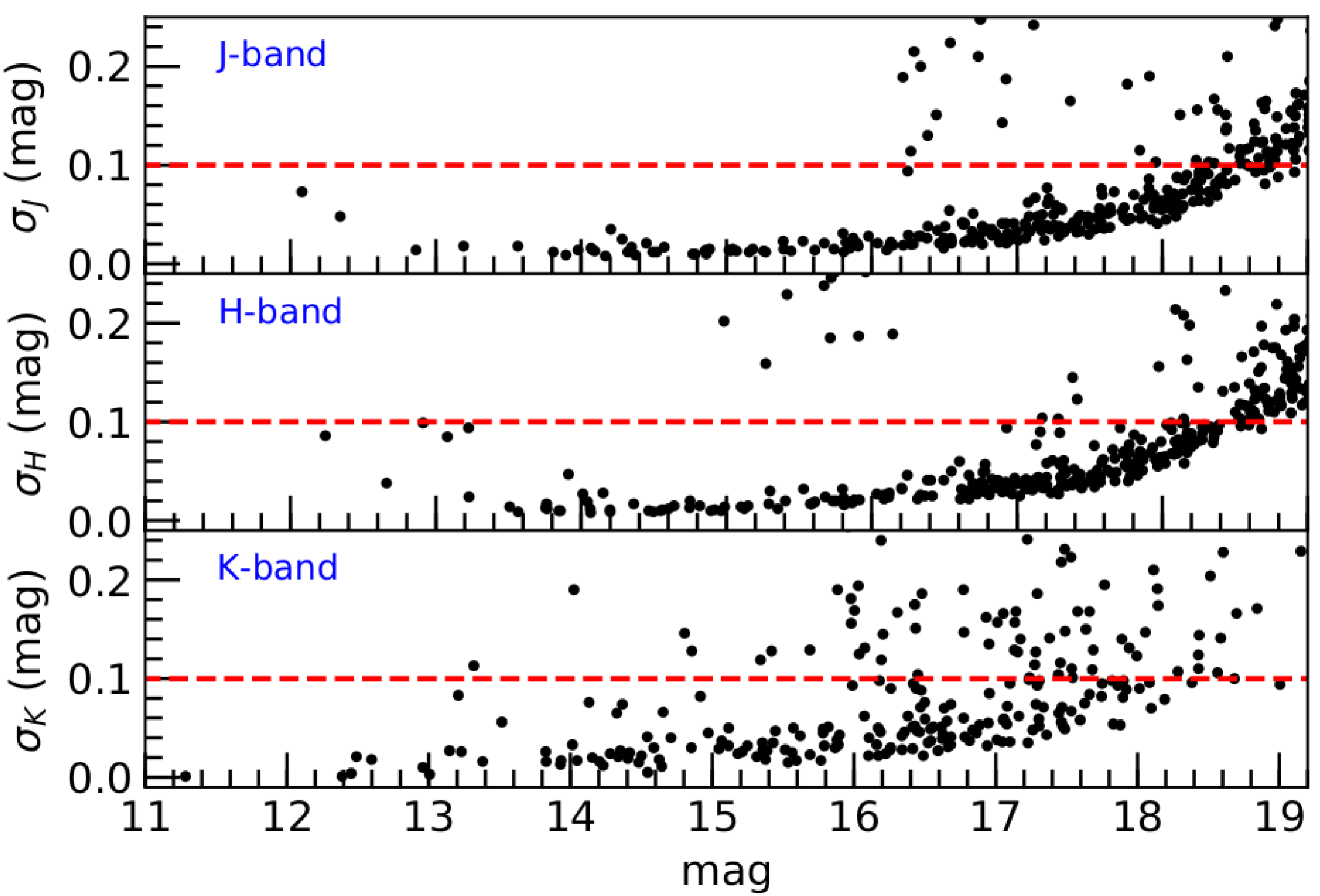}	 
	\caption{TIRCAM2 J, H and K bands magnitudes versus magnitude errors of the globular cluster M53
		observed with an exposure time of 50 s, 50 s and 100 s, respectively (top panel). Aperture photometry was carried out	with an aperture radius of one FWHM. For comparison, J, H and K bands magnitudes versus magnitude
		errors plot of \citet{doi:10.1142/S2251171718500034} is also presented (bottom panel), where the globular cluster M92 was observed with effective exposure times of 550 s, 550 s and 1000s, respectively.}
	\label{Fig: sensitivity_comparison_for_TIRCAM2}
\end{figure}

\subsection{Performance of TIRCAM2 on the side port of DOT}
We observed J, H and K bands images of a Galactic globular cluster, M53 and a galaxy, M99 on May 4, 2020. The standard NIR observation strategy was followed during the data acquisition. Photometric observations in J, H and K bands were taken in five dithered positions, and multiple frames (10) of 10 s exposure were acquired in each dithered position in the case of K band, while in J and H bands, one frame of 50 s exposure was taken in each dithered position. As mentioned above, systematic data acquisition was not possible because of the cloudy weather and highly humid pre-monsoon weather conditions. J band images of M53 and M99 are shown in Fig.~\ref{Fig:Comaparsion_between_2massVsTIRCAM2}. In the case of M53, the mosaic of five dithered J band images is shown which yields a FoV of 2.6 arcmin $\times$ 2.6 arcmin. For comparison, 2MASS J band image of the same region is also presented, and it is evident that the TIRCAM2 image is much deeper and has a better spatial resolution than the 2MASS image. During the night, the full width at half maxima (FWHM) seeing was typically sub-arcsecond ($\sim$ 0.8 arcsec, see Fig.~\ref{Fig: radial_profile_TIRCAM2}) in all J, H and K bands. It is worthwhile to note that the pixel scale of TIRCAM2 is optimized for sub-arcsecond seeing conditions at Devasthal. The J, H and K bands images of M53 are used to estimate the limiting magnitudes of the instrument. We followed the standard NIR photometric reduction techniques to reduce the data with the help of standard tasks of the Image Reduction and Analysis Facility (IRAF\footnote{http://iraf.noao.edu/}, \citealt{Tody1986, Tody1993}). The aperture photometry was carried out using the APPHOT package of IRAF. The zero points of photometry were determined in comparison with 2MASS magnitudes. The 10$\sigma$ limiting magnitudes are about 17.4 mag, 17.2 mag and 17.4 mag in J, H and K bands for an effective exposure time of 50 s, 50 s and 100 s, respectively, as displayed in Fig.~\ref{Fig: sensitivity_comparison_for_TIRCAM2}. Fig.~\ref{Fig: sensitivity_comparison_for_TIRCAM2} also shows the previous measurements of 10$\sigma$ sensitivity limits by \citet{doi:10.1142/S2251171718500034} with an effective exposure of 550 s, 550 s and 1000 s in J, H and K bands, respectively, when it was mounted on the axial port of DOT. Comparison of sensitivity between two measurements is listed in Table~\ref{table:TIRCAM2_sensitivity}. It can be seen that our measurements are 1.6 mag, 1.6 mag and 0.6 mag brighter in J, H and K bands, respectively, than the study of \citet{doi:10.1142/S2251171718500034}. The difference could be because of several factors such as sky brightness, the reflectivity of the primary mirror, effective exposure time and the poor reflectivity of the flip mirror used to direct the light towards the side port instrument and its’ improper cleaning. Furthermore, it can be derived from Pogson's formula that the difference (a factor of 10) between the effective exposure time of two measurements yields a difference in sensitivity limit of $\sim$1.2 mag, however, a few long exposure frames are required to test the performance of the instrument. Further, the spread of magnitudes in our estimation can be seen for the J and H bands, however, no such spreading is apparent for the K band. This could be because of the improper sky subtraction in the case of J and H bands (only 1 frame in each of 5 dithers was taken as mentioned earlier). Thus, in the upcoming DOT cycle, we plan to obtain systematic data for calibration as well as for deriving colour transformation equations.

\begin{table}[htb]
	%% use tabular font for a smaller size font
	\tabularfont
	\caption{Comparison of 10$\sigma$ sensitivity limits between TIRCAM2 on the side port and axial port of DOT.}\label{table:TIRCAM2_sensitivity} %%10/12
	\begin{tabular}{l|crc|crc}
		\topline
		Filter& Mag* &Exp (s) & Ref & Mag*&Exp (s)&Ref\\\midline
		J& 17.4 & 50 & 1&  19.0 & 550 & 2\\
		H& 17.2 & 50 & 1 &18.8  & 550 & 2\\
		K& 17.4 & 100 & 1& 18.0 & 1000 & 2\\
		\hline
	\end{tabular}
	%%use \tablenotes{footnote} to get the table foot note
	\tablenotes{* 10$\sigma$ limiting magnitude\\
		 1 $-$ This work (side port), 2 $-$ \citet{doi:10.1142/S2251171718500034} (axial port)}%%9/11
\end{table}

\section{TIFR Near Infrared Spectrometer and Imager (TIRSPEC)} \label{TIRSPEC}
The TIRSPEC was designed and built-in in collaboration with MKIR. It offers spectroscopy and imaging modes of observations with various filters covering 1 to 2.5 $\mu$m wavelength regime. The TIRSPEC spectroscopic mode includes two modes, single order and cross-dispersed, with slit widths ranging from 1 to 7.92 arcsec. In the single order mode, slits of 300 arcsec length are used with order sorter filters in Y, J, H and K bands to obtain the spectrum in each order separately, while in the cross-dispersed mode, a grism is used as the dispersing element with slit length of 10 arcsec to provide a resolving power, R of 1200. Two cross-dispersed modes, YJ and HK, are used for observations to get larger wavelength coverage in one exposure than single order mode. The YJ mode covers Y and J windows (1.02--1.49 $\mu$m) and the HK mode covers H and K windows (1.50--1.84 $\mu$m and 1.95--2.45 $\mu$m). For imaging observations, the instrument provides broad band J, H, Ks and narrow band Methane (on and off), [Fe {\footnotesize II}], H2,  BrG, K-Cont and CO filters.  The TIRSPEC employs a Teledyne 1024 $\times$ 1024 pixel Hawaii-1 PACE array detector with a cutoff wavelength of 2.5 $\mu$m and it covers a FoV of 307 arcsec $\times$ 307 arcsec with an image scale of 0.3 arcsec/pixel on 2 m HCT. The TIRSPEC was mounted successfully on the side port of the HCT during June 2013 for the engineering and scientific runs on the telescope, and the instrument saw the first light on June 21, 2013. The subsequent characterization and astronomical observations were carried out during July and August 2013. Over the next few months, minor mechanical modifications in the filter movement mechanisms to improve the movement of the filter wheels, the up-gradation of slit lengths to 50 arcsec, and the optimization of various slits' positions for efficient observations were made. The TIRSPEC was commissioned successfully and the subsequent characterization and astronomical observations were completed. The TIRSPEC was made available to the worldwide astronomical community for science observations since May 2014. Subsequently, sub-array readout capability has been incorporated into the system to allow for photometry of brighter objects. For additional details, we refer \citet{2012ASInC...4..191O} and \citet{2014JAI.....350006}. The TIRSPEC Science Case is broad and includes various studies in star formation (e.g., \citealt{2015ApJ...815....4N,2016ApJ...833...85B}), young M-dwarfs \citep{2020MNRAS.493.4533K}, K--M giants \citep{2019MNRAS.484.4619G}, asymptotic giant branch stars (e.g., \citealt{2018aj..155..216, 2021AJ....161..198G, 2021MNRAS.506.1962S}) and supernova \citep{2016MNRAS.457.1000S}. Currently, about 50\% of the observing proposals on HCT use TIRSPEC as a focal plane instrument. Recently, long-duration exposure for faint source spectroscopy has been introduced in TIRSPEC and initial analysis shows this seems to be giving a better S/N ratio. The TIRSPEC instrument is now proposed to be upgraded to use a state-of-the-art NIR array detector (1K $\times$ 1K H1RG).

\section{TIFR-ARIES Near Infrared Spectrometer (TANSPEC)} \label{TANSPEC}
The TANSPEC is a NIR spectrometer that has been developed in collaboration with TIFR, ARIES and MKIR for the DOT. It offers various modes of observations which include spectroscopy with slits of different lengths and widths as well as imaging with broad and narrow band filters. Importantly, the uniqueness of the instrument lies in its capability of simultaneous wavelength coverage ranging from 0.5 to 2.5 $\mu$m in spectroscopic mode. In TANSPEC, reflective slit mirrors are used to split the incoming light beam into two (see the overall optical layout of TANSPEC in Fig. 6 of \citealt{2018BSRSL..87...58O}). A portion of the light beam is transmitted to the spectrometer and the remaining light is reflected to the slit viewer camera which acts as the telescope guider (infrared guider) and pupil viewer for instrument alignment on the telescope as well as NIR imager. Thus, the TANSPEC uses two detectors -- one is 2K $\times$ 2K H2RG array for spectroscopy, and the other is 1K $\times$ 1K H1RG array with a FoV of 1 arcmin $\times$ 1 arcmin and a plate scale of 0.25 arcsec/pixel for imaging. The detectors are cooled to 76 $\pm$ 2 K during operation through hybrid system with a closed-cycle Helium cryo-cooler and liquid Nitrogen. The TANSPEC houses a range of slits of widths and lengths from 0.5 arcsec to 4.0 arcsec and from 20 arcsec to 60 arcsec, respectively. It provides two modes of spectroscopic observations, prism (LR) mode and cross-dispersed (XD) mode with resolutions (R) of $\sim$ 100--350 and 2750, respectively, in the narrowest available slit of width 0.5 arcsec. In XD mode, a combination of grating and two cross-dispersing prisms are used to pack all the 10 orders (order 3 to order 12) onto the detector. Slit viewer camera consists of broad band r', i', Y, J, H, Ks and narrow band H2 \& BrG filters for imaging observations. The TANSPEC can be used for a wide range of scientific research from local star formation to extragalactic astronomy. The LR mode is efficient for long-term survey programs due to its high throughput. The TANSPEC is one of the few worldwide instruments at present that enables simultaneous observations in optical and NIR. It is operational from the axial port of DOT during its scheduled allocation. Additional technical details of the instrument will be presented in Sharma {\em et al.} (in preparation).

%\begin{figure}
%	\centering\includegraphics[height=.3\textheight]{TANSPEC_optical_layout.pdf}
%	\caption{Overall optical layout of TANSPEC.}
%	\label{Fig:TANSPEC_optical_layout}
%\end{figure} 

\subsection{Performance of TANSPEC on DOT}
\begin{figure}
	\centering\includegraphics[height=.25\textheight]{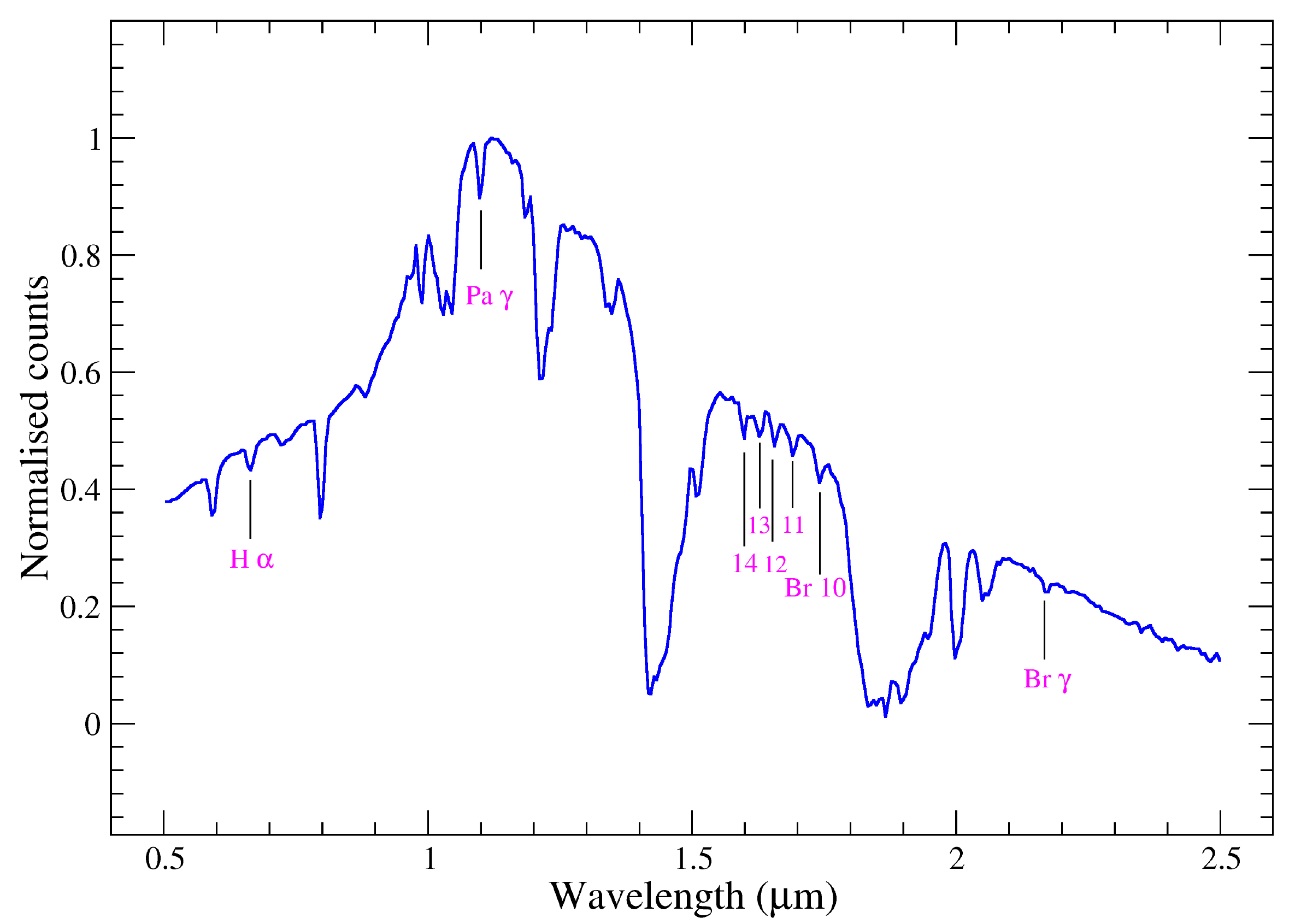}
	\caption{Wavelength calibrated TANSPEC spectrum of an A-type star in LR mode (observed with 0.5 arcsec slit). The spectrum is normalised at 1.1 $\mu$m. Most of the unmarked features are due to terrestrial atmospheric O$_2$ and H$_2$O absorption.}
	\label{Fig:TANSPEC-LR-spectrum}
\end{figure}

\begin{figure*}
	\centering\includegraphics[height=.55\textheight]{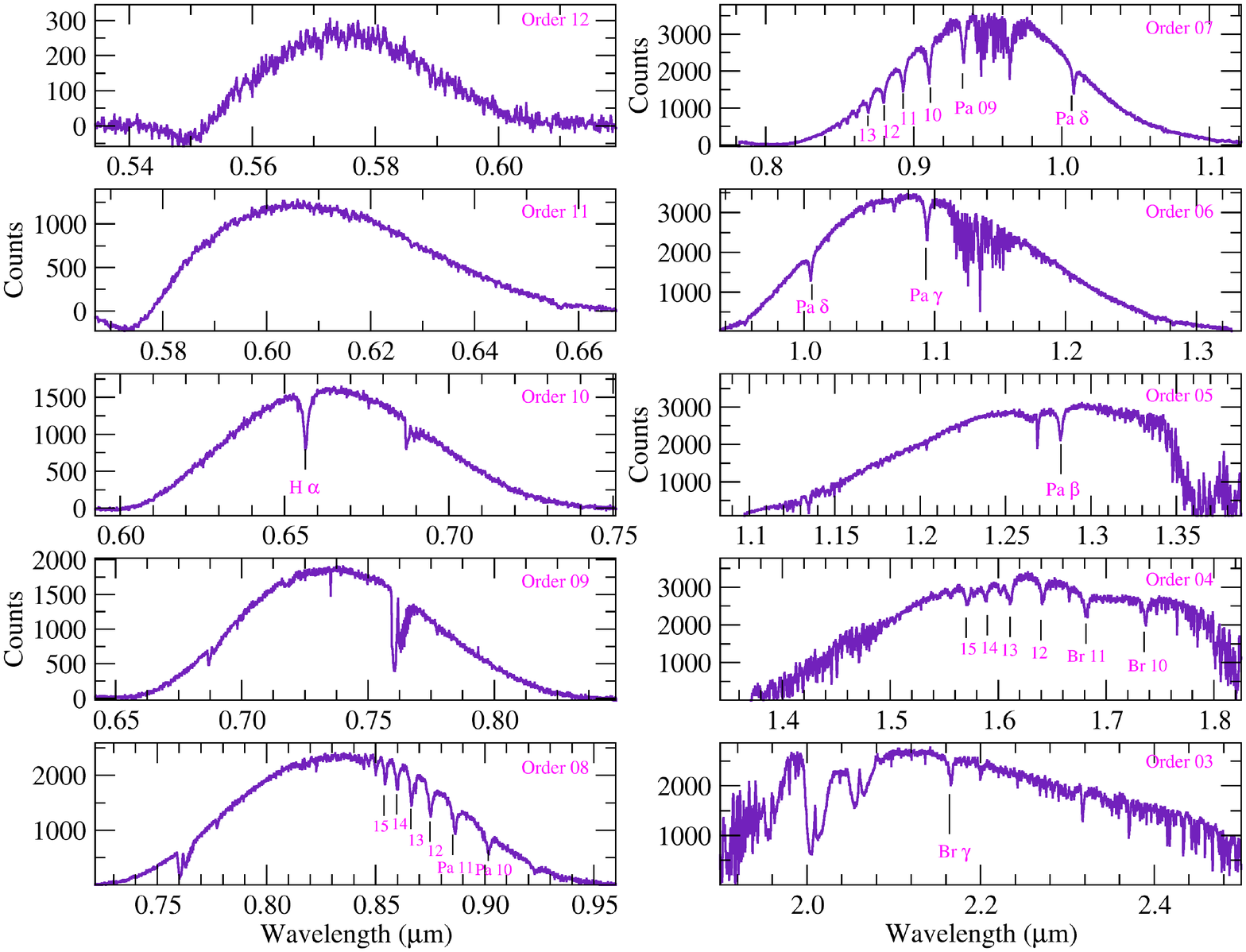}
	\caption{Wavelength calibrated TANSPEC spectra of an A-type star in XD mode (observed with 0.5 arcsec slit). It is to note that the spectra cover all the orders and provide the wavelength coverage ranging from 0.5 to 2.5 $\mu$m. Most of the unmarked features are due to terrestrial atmospheric O$_2$ and H$_2$O absorption.}
	\label{Fig:TANSPEC-XD-spectrum}
\end{figure*}

The TANSPEC was mounted successfully on the axial port of the DOT on April 2, 2019 and the first light took place on April 12, 2019. The characterization of the TANSPEC was performed over the next few DOT cycles. Science observations of several astronomical sources were also carried out during this time. After successful commissioning and the subsequent characterization, it was seen that the instrument performed as expected. Thus, the TANSPEC was made available to Indian and Belgian astronomers since October 2020. In the following, we give a brief outline of the instrument performance. The spectral resolution in each order was estimated by measuring the FWHM of argon and neon lamps lines through the slits of 0.5 arcsec and 1.0 arcsec width. The measured resolutions are 2800 and 1700 on average for  0.5 arcsec and 1.0 arcsec slits, respectively, in XD mode. For LR mode, the values are 100 -- 400 and 20 -- 100 for 0.5 arcsec and 1.0 arcsec slits, respectively. Attainable spectroscopic sensitivity for different magnitudes was estimated by taking spectra of various stars. We have found that the required exposure times for continuum S/N $\sim$ 10 in the J band of a star (J = 14 mag) are 1000 s and 100 s for  0.5 arcsec and 1.0 arcsec slits in XD mode, respectively, in typical DOT night conditions. We also found that the 10$\sigma$ limiting magnitudes in 1 minute are about 18.5 mag, 17.7 mag and 17.2 mag in J, H and K bands, respectively. For example, wavelength calibrated spectra of an A-type star are displayed for LR and XD modes in Fig.~\ref{Fig:TANSPEC-LR-spectrum} and Fig.~\ref{Fig:TANSPEC-XD-spectrum}, respectively. The detailed characterization of the instrument will be presented in Sharma {\em et al.} (in preparation). 

%\begin{figure}
%	\centering\includegraphics[height=.25\textheight]{m53-tanspec.pdf}
%	\caption{Image of a Galactic globular cluster M53 taken in the K band using the slit-viewer of TANSPEC on 3.6 m DOT on April 20, 2019.}
%	\label{Fig:TANSPEC-image}
%\end{figure}

\section{Far-infrared observations using TIFR 100-cm balloon-borne telescope} \label{FPS100}
Balloon-borne FIR astronomy is a part of the TIFR-Japan collaboration project where a Japanese (ISAS-JAXA and Nagoya University, Japan) FPS is used at the focal plane of the T100. The primary motivation to develop such a facility is to probe the GSFRs through [C {\footnotesize II}]  intensity map at 157.74 $\mu$m as it is the most important gas coolant in low-density photo-dissociation regions formed around an H {\footnotesize II} region by massive stars. FPS consists of a pair of Fabry-Perot interferometers, the Scanning Fabry-Perot (SFP) and the Fixed Fabry-Perot (FFP). Each interferometer consists of etalon plates which are composed of nickel mesh, suitable for FIR observations. The SFP with moving etalons has an order of $>$ 70 and a finesse of $\sim$ 26 and provides a finite spectral range (157.41 -- 158.57 $\mu$m). On the other hand, the FFP with fixed etalon is tuned to forbidden fine structure [C {\footnotesize II}] line at 157.74 $\mu$m. It has an order of only 3 and acts as an order sorting filter allowing only the desired wavelengths to pass through it to the single stressed Ge:Ga detector. The detector is cooled to 1.8 K using liquid $^4$He under reduced surface pressure. T100 provides a spatial resolution of $\sim$ 1.0 arcmin at $\sim$ 158 $\mu$m and FPS has spectral resolution of R $\sim$ 1800 ($\sim$ 175 km/s). Balloons are launched from the TIFR Balloon Facility at Hyderabad during two seasons in a year: summer (January - April) and winter (October - December).

FPS100 was first launched in 1999 and latest on October 28, 2018.  From February, 2017 to October, 2018, the FPS100 payload was successfully flown four times, and it mapped large regions in [C {\footnotesize II}] line and continuum of several GSFRs such as RCW 38, Orion Bar, W3, RCW 36, NGC 2024, W31, Carina, NGC 6334 and NGC 6357 in the fast scanning mode. 

The up-gradation of T100 electronics to incorporate a new Japanese 5 $\times$ 5 FPS array is undergoing. T100 with new FPS will be efficient to map large-area with high-spatial ($\leq$ 40 arcsec) and spectral (R $\sim$ 10000; $\sim$ 30 km/s) resolutions. It will then allow the investigation of gas geometry and dynamics of filaments in GSFRs in unprecedented detail. The first light of the upgraded FPS100 is planned for the near future.

\section{Infra-Red Spectroscopic Imaging Survey (IRSIS) Payload} \label{IRSIS}
The IRSIS payload was planned for an Indian Satellite aiming to carry out an infrared spectroscopic imaging survey in the wavelength regime from 1.7 to 6.4 $\mu$m. A medium-size Ritchey-Chretien telescope with a paraboloid primary mirror of aperture $\sim$ 300 mm and a secondary mirror of aperture $\sim$ 36 mm will collect infrared light from astronomical objects and focus and feed the incoming light into an array of micro-lenses coupled to infrared optical-fibre bundles at the Cassegrain focus of the telescope. The other end of the fibre bundles is rearranged into multiple slits of the spectrometer. The spectrometer has two channels: the Short-Wavelength Channel covers 1.7 $-$ 3.4 $\mu$m and the Long-Wavelength Channel covers 3.2 $-$ 6.4 $\mu$m, and each Channel of the spectrometer uses transmission gratings to disperse images of the multiple slits onto a 1K $\times$ 1K HgCdTe detector array (cooled at 80 K). Thus, two alike detector arrays in both the Channels simultaneously capture the spectra and provide seamless instantaneous coverage over the entire wavelength range of 1.7 $-$ 6.4 $\mu$m.

The optomechanical design of the Laboratory Model (LM) of the payload has been finalized. The facilities in TIFR central workshop have been used to fabricate the mechanical parts. The infrared optical fibres and the Focal-Ratio-Degradation characteristics of these fibres were assessed in the laboratory.  The lenslets have been modelled and procured. The test integral field unit consisting of 70 $\mu$m core optic fibres was assembled and probed successfully. The large-format (2K $\times$ 2K) functional grade H2RG infrared detector array and a special read-out and front-end signal processing ASIC are already verified. The detector characterization (quantum efficiency, noise and pixel operability) has been performed in the laboratory. The detector was also probed in different modes by lamp spectra in the laboratory.  A clean room facility has been set up within the available laboratory space of the IRA group at TIFR to be self-sufficient in routine assembly and testing of sensitive components of IRSIS. In-house modelling of the LM of IRSIS payload was accomplished at TIFR.  The final Laboratory Phase Model report has been submitted to ISRO (Space Science Office).
%%Use table* environment to get the table spanning both the columns

%\begin{table*}[htb]
%\tabularfont
%\caption{Caption text here}\label{secondTable}
%\begin{tabular}{lccccccccccccr}
%\topline
%\textbf{head1}&\multicolumn{11}{c}{\textbf{head2}}&\textbf{head3}\\
%\midline
%one& two &three&four&five&six&seven&eight&nine&ten&eleven&twelve&thirteen\\
%1&2&3&4&5&6&7&8&9&10&11&12&13\\
%aaa&bbbb&cccc&ddddd&eee&ffff&ggggg&hhhhhhhh&iiii&kkkkkk&hhh&jjjjjj&lllll\\
%\hline
%\end{tabular}
%\tablenotes{Table footnote here. Table spanning both the columns.}
%\end{table*}

%%An example of a figure

%\begin{figure}[!t]
%\includegraphics[width=.8\columnwidth]{fig1.eps}
%\caption{caption goes here}\label{figOne}
%\end{figure}

%%An example of a double column figure
%%Use figure* environment

%\begin{figure*}
%\centering\includegraphics[height=.15\textheight]{fig1.eps}
%\caption{caption spanning two columns}
%\centering\includegraphics[height=.25\textheight]{fig1.eps}
%\caption{caption here}
%\end{figure*}

%\vspace{-4em}
\section{Conclusion}
The TIRSPEC on the 2 m HCT with imaging and spectroscopy capabilities in the wavelength regime ranging from 1 to 2.5 $\mu$m, the TANSPEC on the 3.6 m DOT with simultaneous spectroscopy capability from 0.5 to 2.5 $\mu$m as well as imaging capability and the TIRCAM2 on the DOT with imaging capability from 1 to 3.7 $\mu$m are the main workhorse for Indian NIR astronomy. These instruments are extremely sensitive to cool stellar photospheres (T $\leq$ 2500 K) and objects surrounded by warm dust envelopes or embedded in dust/molecular clouds, and hence, are suitable for a variety of challenging astrophysical problems, from very low mass stellar populations (M dwarfs, brown dwarfs) to strong mass-losing stars on the AGB and from young stellar objects still in their protostellar envelopes to active galactic nuclei. One can avail of these facilities through standard proposal submission subject to the scientific merit of the proposals. FIR observation is also possible through TIFR balloon-borne telescope to map [C {\footnotesize II}] line at 157.74 $\mu$m in the GSFRs, and with the up-gradation of balloon-borne telescope with 5 $\times$ 5 FPS array, it will then be possible to investigate gas geometry and dynamics of filaments in the GSFRs with high-spatial and spectral region covering large-area. In addition, dedicated survey facilities such as IRSIS to carry out the spectroscopic imaging survey in the wavelength regime from 1.7 to 6.4 $\mu$m is underway. With all these existing and upcoming facilities Indian observation astronomy is heading towards a bright future and TIFR is happy to serve the astronomical community.

%%Appendix5 × 5 FPS array

\appendix

%\section{An appendix section}
%Text goes here (Radhakrishnan 1980).
%\begin{equation}
%x=a+b+c
%\end{equation}

%%\section{Another appendix section}
%Text goes here.
%\begin{equation}
%y^2=ax+b+c
%\end{equation}
%\vspace{-3em}

%%Use section* for acknowledgements
\section*{Acknowledgements}
The authors are very much thankful to the reviewer for their critical and valuable comments, which helped us to improve the paper. This work is supported by the Tata Institute of Fundamental Research, Mumbai under the Department of Atomic Energy, Government of India. SG, DKO and MBN acknowledge the support of the Department of Atomic Energy, Government of India, under project Identification No. RTI 4002. We also thank all the members of the IRA group at TIFR for their support during observations. The authors would like to thank the staff at the 3.6 m DOT, Devasthal and ARIES, for their cooperation during the installation and characterization of TIRCAM2 and TANSPEC and the staff of IAO, Hanle and CREST, Hosakote, that made TIRSPEC observations possible. The authors are thankful to Mr. Douglas W. Toomey and the entire MKIR team for their contribution to the TIRSPEC and TANSPEC projects.

\vspace{-1em}

%%use \balance somewhere in the left column of the last page to balance the two columns in the end page

%%References section
%\begin{theunbibliography}{}
%\vspace{-1.5em}

%\bibitem{latexcompanion}
%Clark D. H., Caswell J. L. 1976, MNRAS, 174, 267
%\bibitem{latexcompanion}
%Dickey, J. M., Salpeter, E. E., Terzian, Y. 1978, Astrophys. J. Suppl. Ser., 36, 77
%\bibitem{latexcompanion}
%Radhakrishnan, G. C. {\em et al.} 1980, in Evans A., Bode M. F., eds, Non-Solar Gamma Rays (COSPAR), Pergamon Press, %Oxford, p. 163
%\bibitem{latexcompanion}
%Starrfield S., Iliadis C., Hix W. R. 2008, in Bode M. F., Evans A., eds, Classical Novae, 2nd edition, Cambridge %University Press, Cambridge, p. 77
%\bibitem{latexcompanion}
%Van Loon J. Th. 2008, in Evans A. et al., eds, R S Ophiuchi (2006) and the Recurrent Nova Phenomenon, ASP Conference %Series, Volume 401, p. 90
%\bibitem{latexcompanion}
%Zwicky, F. 1957, Morphological Astronomy, Springer-Verlag, Berlin, p. 258
%

%\end{theunbibliography}
%\bibliographystyle{aasjournal}
\bibliography{reference}

%width=9cm,height=5cm,keepaspectratio
%2MASS vs TIRCAM2 image comparison
% \begin{table*} 
%	\begin{tabular}{cc} 
%		\includegraphics[scale = 0.95]{M53J-TIRCAM2.pdf} & \includegraphics[scale = 1.0]{M53-2MASS-b.pdf} \\
%		\includegraphics[width=3.8cm,height=3.2cm]{M99J-TIRCAM2-1.pdf} & \includegraphics[scale = 0.84]{M99-2MASS.pdf} \\
%	\end{tabular} 
%	\caption{...blaaaaaaaaaa blaaaaaaaaaaaaaaaaaaa blaaaaaaaaaaaaaaaaaa blaaaaaaaaaaaa} 
%	\label{Figure:Hierarchical_Bayesian_Linear_model_fit} 
% \end{table*} 
%\end{center}

%sensitivity comaprison of TIRCAM2 on DOT axial port and side port
%\begin{center} 
%	\begin{table*} 
%		\begin{tabular}{cc} 
%			\includegraphics[scale = 0.25]{Tapas2017_Tircam2_sensitivity.pdf} & \includegraphics[scale = 0.5]{m53-j.2nd.pdf}  \\
%		\end{tabular} 
%		\caption{...blaaaaaaaaaa blaaaaaaaaaaaaaaaaaaa blaaaaaaaaaaaaaaaaaa blaaaaaaaaaaaa} 
%		\label{Figure:Hierarchical_Bayesian_Linear_model_fit} 
%	\end{table*} 
%\end{center}
%\begin{center} 
%	\begin{table*} 
%		\begin{tabular}{cc} 
%			\includegraphics[width=9cm,height=5cm,keepaspectratio]{M53-2MASS-b.pdf} & \includegraphics[width=9cm,height=5cm,keepaspectratio]{M53J-TIRCAM2.pdf}  \\
%			\includegraphics[width=6cm,height=5cm,keepaspec]{M99-2MASS.pdf} & \includegraphics[width=9cm,height=5cm,keepaspectratio]{M99J-TIRCAM2-1.pdf}  \\
%		\end{tabular} 
%		\caption{...} 
%		\label{Figure:Hierarchical_Bayesian_Linear_model_fit} 
%	\end{table*} 
%\end{center}

\end{document}